 \documentclass[pmlr,twocolumn,10pt]{jmlr} 

\usepackage{listings}

\usepackage{booktabs}
\usepackage{siunitx}

\usepackage[switch]{lineno}

\usepackage{subcaption}
\usepackage{hyperref}

\theorembodyfont{\upshape}
\theoremheaderfont{\scshape}
\theorempostheader{:}
\theoremsep{\newline}

\jmlrworkshop{Machine Learning for Health (ML4H) 2025} 

\title[FDA AI Search]{FDA AI Search: Making FDA-Authorized AI Devices Searchable}

\author{%
\Name{Arun Kavishwar}\Email{arun\_kavishwar@alumni.brown.edu}\\
\addr Dana-Farber Cancer Institute\\
\AND
\Name{William Lotter} \Email{lotterb@ds.dfci.harvard.edu}\\
\addr Dana-Farber Cancer Institute, Brigham and Women's Hospital, \& Harvard Medical School
}

\begin{document}

\maketitle

\begin{abstract}

Over 1,200 AI-enabled medical devices have received marketing authorization from the U.S. FDA, yet identifying devices suited to specific clinical needs remains challenging because the FDA’s databases contain only limited metadata and non-searchable summary PDFs. To address this gap, we developed \textbf{FDA AI Search}, a website that enables semantic querying of FDA-authorized AI-enabled devices. The backend includes an embedding-based retrieval system, where LLM-extracted features from authorization summaries are compared to user queries to find relevant matches. We present quantitative and qualitative evaluation that support the effectiveness of the retrieval algorithm compared to keyword-based methods. As FDA-authorized AI devices become increasingly prevalent and their use cases expand, we envision that the tool will assist healthcare providers in identifying devices aligned with their clinical needs and support developers in formulating novel AI applications.

\end{abstract}

\paragraph*{Data and Code Availability}
The data used is sourced from the FDA's database of AI-enabled devices\footnote{\url{https://www.fda.gov/medical-devices/software-medical-device-samd/artificial-intelligence-enabled-medical-devices}} and accompanying authorization summary documents. The database version used corresponds to an update on July 10th, 2025. Code is available at \url{https://github.com/lotterlab/fda_ai_search}.

\paragraph*{Institutional Review Board (IRB)}
This research does not require IRB approval.

\section{Introduction}
\label{sec:intro}

Over 1,200 AI-enabled medical devices have received marketing authorization from the U.S. Food and Drug Administration (FDA). Realizing their collective potential requires connecting the specific needs of healthcare providers with devices designed to address them. This remains difficult, as the FDA's database of AI-enabled devices only contains limited metadata (e.g., company name, review panel) and links to non-searchable PDF summaries. While recent efforts have categorized devices along dimensions such as input data type \citep{Singh2025-er} or regulatory path \citep{Muehlematter2023-st} or focus on specific use cases \citep{Ebrahimian2022-xw, McNamara2024-ci, Milam2023-ma, Morey2025-ik}, a more flexible and user-friendly solution is needed.

To address this gap, we developed \textbf{FDA AI Search}, a website that enables semantic querying of FDA-authorized AI-enabled devices. Using the public summary statements that accompany FDA marketing authorization, we generate text embeddings for each device and retrieve relevant matches to user queries based on embedding similarity. In addition to introducing the tool and its underlying methodology, we present quantitative and qualitative evaluations compared to traditional keyword-based search methods.

\section{Methods}
\begin{figure*}[t]
    \centering
    \includegraphics[width=\linewidth]{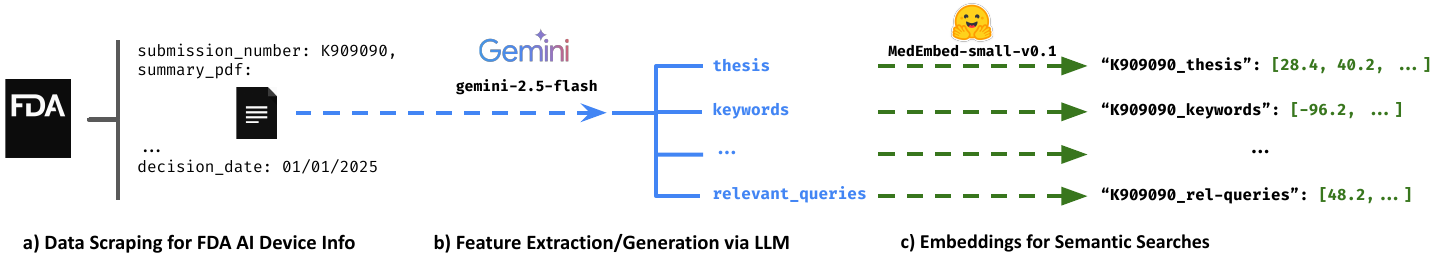}
    \caption{Embedding generation process. a) FDA databases are scraped to retrieve metadata and the summary PDF for each AI device. b) Gemini-2.5-flash is prompted to generate a set of text features from the summary PDF and metadata. c) Each feature is embedded using MedEmbed for semantic search.}
    \label{fig:highlevelflow}
\end{figure*}

The backend of the website consists of an embedding-based retrieval system. Embeddings were generated for each of the 1,247 AI-enabled devices listed by the FDA as of July 10th, 2025 (Figure~\ref{fig:highlevelflow}). These embeddings were generated from two sources: 1) basic metadata from the FDA's AI-enabled device database and 2) the public summary accompanying FDA authorization, as described below.

\subsection{Data Curation}
The FDA’s list of AI-enabled devices provides basic tabular information for each marketing authorization along with a link to the corresponding entry in the FDA’s 510(k), De Novo, or PMA databases, depending on the regulatory pathway. Each database entry includes a link to the authorization summary PDF, which we retrieved for every device. In rare cases ($\sim$1\%), the entire submission (with proprietary information redacted) was also released under the Freedom of Information Act, in which case we used that document rather than the summary alone.

\subsection{Feature Extraction}
Rather than simply creating an embedding for each device in a single shot, embeddings were created in a structured fashion to focus on information likely relevant to user queries. For each device PDF, Gemini-2.5-flash \citep{gemini} was initially prompted to extract the following five features: `summary', `keywords', `relevant questions', `thesis' (i.e., concise statement of purpose), and `key concepts'. The full prompt and further details of this process are provided in the Appendix. Additional features were then generated from the extracted features to help further align with potential queries. The first generated feature was a `search boost' term consisting of the company name concatenated with the device name and extracted keywords, where the company and device names were extracted from the FDA's AI database. The other generated features consisted of `query match' terms wherein Gemini-2.5-flash was prompted to generate three distinct search queries that a clinician would use to find this device based on the originally extracted set of five features (see also Appendix). Of the 9 total features (5 original features and 4 generated features), 7 were ultimately chosen to be included in embedding generation: `keywords', `relevant questions', `thesis', `search boost', `query match 1', `query match 2', `query match 3'. The `summary' and `key concepts' features were excluded because they were found to be redundant with `thesis' and `keywords', respectively.

\subsection{Generating Embeddings}
Embeddings for each of the 7 features for each device were generated using MedEmbed \citep{medembed}, an NLP embedding model specifically trained for medical and clinical data. The small version of MedEmbed was used (MedEmbed-small-v0.1), as it was the only one that created vectors small enough to host with the designed frontend deployment (384 dimensions / vector). This version is also used when creating embeddings of user queries for the retrieval process.

\subsection{Retrieval}

For query $q$, a device ($d$) score was calculated based on the weighted sum of the cosine similarity between the embedded query ($e_q$) and the feature embedding vectors $e_d^i$, along with a standard bag-of-words keyword scoring function, BM25 (Equation~\ref{eq:score}). BM25 is a common ranking function used by search engines \citep{bm25} and was included to complement the AI-based semantic similarity. The BM25 input for a device was a concatenation of the keywords, relevant questions, thesis, key concepts, and search boost features. The embedding weights ($w_i$) in Eq.~\ref{eq:score} were calculated via a Bayesian optimization search and the $\lambda$ parameter controlling the trade-off between the embedding similarity and BM25 was calculated via a grid search, as detailed below.

\begin{align}
\operatorname{Score}(q,d)
&= \frac{\lambda}{\sum_{i=1}^{n=7} w_i}
   \sum_{i=1}^{n=7} w_i \,\mathrm{sim}\!\left(e_q, e_d^i\right) \notag \\
&\quad + (1-\lambda)\,\operatorname{BM25}(q,d)
\label{eq:score}
\end{align}

\subsubsection{Optimizing Retrieval Weights}
\label{subsec:weights}
For the embedding weights optimization, a simulated validation set was created by randomly sampling 50 devices and generating a hypothetical query for each device. This query was generated using Google's Gemma-3 (specifically, gemma3n:e2b, a 2B parameter model designed to be computationally efficient). The model was prompted to create a clinically relevant search query based on the thesis and keywords of each device (see Appendix).

The weights for the 7 features were optimized using Optuna \citep{optuna}, a hyperparameter optimization software package. The objective function for optimization was the mean Hit Rate@K with K=5 across the 50 sampled devices (if the ground truth device was in the top K=5 returned results, the retrieval was considered a hit). The weights were optimized over a number of trials, where for each trial, Optuna samples a set of weights (constrained to the range of [0.01, 0.5]), the objective function score is recorded, and then new weights are sampled for the next trial based on the cumulative trial results. The default Tree-structured Parzen Estimator sampler was used. After each trial, 20\% of the device-query pairs were randomly replaced to avoid overfitting.

Over 1830 trials, the set of weights scoring highest on the objective function was recorded. Once these weights had been determined, the only hyperparameter remaining was $\lambda$, the relative weighting between the embedding search and the bag-of-words metric. Using the same methodology of creating test cases, a grid search was performed over 21 incremental values of $\lambda$ with the frozen embedding weights. The final values of the embeddings weights ($w_i$) and $\lambda$ are contained in the Appendix.

\subsection{User Interface}
\begin{figure}[htbp]
\centering
\includegraphics[width=1.0\linewidth]
{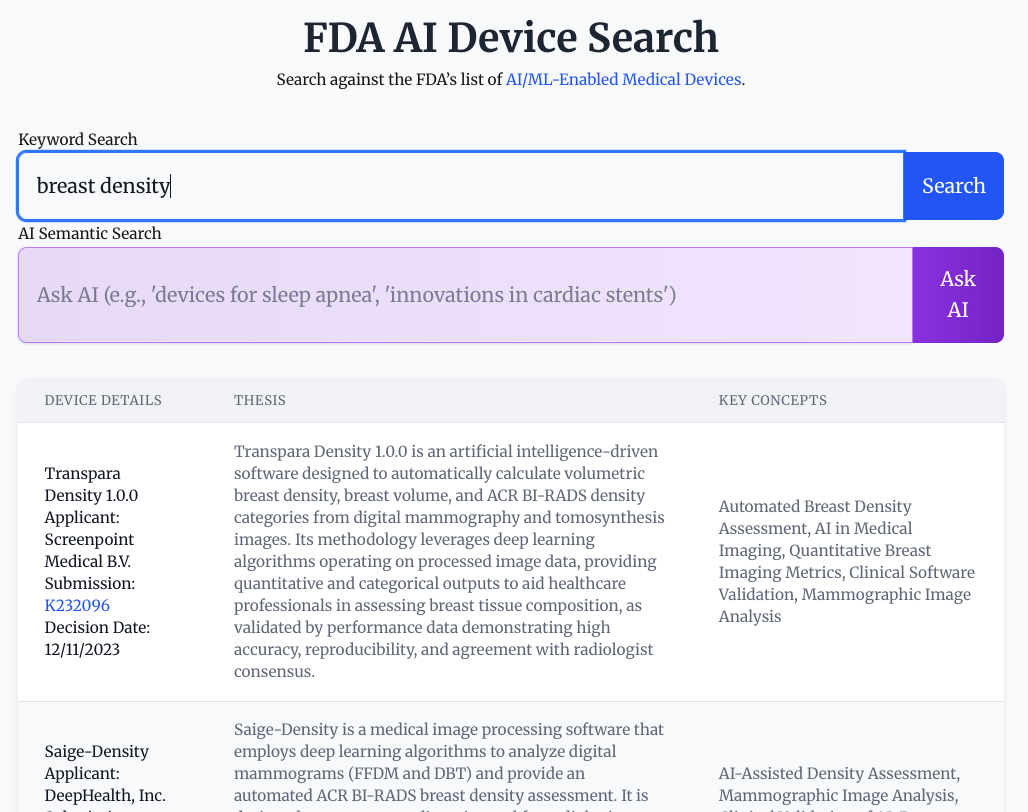}
\caption{Sample UI view and keyword search.} \label{fig:sample-ui}
\end{figure}

The code was written in React/Next.js and deployed via Vercel. The website is hosted at \url{fda-ai-search.com}.
For thoroughness and user experience, a keyword search option is also included that performs a simple lookup of the query against concatenated text for the device: the submission number, device name, company, thesis, keywords, and key concepts. If all the words in the query are found in this concatenated text snippet, the device is presented as an option. This search function can be useful for fast lookup queries, as illustrated in Figure \ref{fig:sample-ui}.

\vspace{-20pt}
\section{Results}
\subsection{Quantitative Evaluation}
While the current lack of curated device information makes evaluation challenging, a pilot quantitative assessment of the retrieval scoring system was performed using a curated subset of FDA AI-enabled devices. Specifically, the FDA AI CAD database from \citet{McNamara2024-ci} was used, which contains manually curated information for 140 FDA-authorized devices designed for medical image interpretation. For each device, a query was created based on the indicated disease for the device (e.g., `lung cancer'), and separately by combining the disease with the indicated data modality (e.g., `lung cancer CT'), as searching for devices related to certain diseases and modalities represents a promising use case for healthcare providers in clinical practice.

Results were measured for the embedding search alone, the BM25 search alone, and the full hybrid search. For a given query, each of these variations results in a ranked list of the 140 devices. The position of the ground-truth device (i.e., the device used to create the query) within this ranked list was recorded and used to quantify performance across the 140 queries. Metrics included summary statistics of the ground-truth position and the Hit Rate@K, measuring the proportion of times the ground-truth device fell within the top K $\in$ \{1, 3, 5, 10\} positions. When a query matched more than one device (e.g., two devices have the same indicated disease), the position of the first match in the ranked list was used.

\begin{table}[h]
\centering
\begin{tabular}{lccc}
\hline
Metric & Embedding & BM25 & Hybrid \\
\hline
Average position & 1.54 & 1.59 & \textbf{1.37} \\
Median position  & \textbf{1.00} & \textbf{1.00} & \textbf{1.00} \\
Min position     & \textbf{1.00} & \textbf{1.00} & \textbf{1.00} \\
Max position     & 8.00 & \textbf{7.00} & \textbf{7.00} \\
Stdev position           & 1.36 & 1.23 & \textbf{1.12} \\
Hit@K = 1        & 0.756 & 0.707 & \textbf{0.829} \\
Hit@K = 3        & 0.951 & 0.927 & \textbf{0.951} \\
Hit@K = 5        & 0.951 & \textbf{0.976} & \textbf{0.976} \\
Hit@K = 10       & \textbf{1.00} & \textbf{1.00} & \textbf{1.00} \\
\hline
\end{tabular}
\vspace{-5pt}
\caption{Disease + Modality Query Evaluation}
\label{table:disease-modality}
\end{table}
\vspace{-15pt}
\begin{table}[h]
\centering
\begin{tabular}{lccc}
\hline
Metric & Embedding & BM25 & Hybrid \\
\hline
Average position & 2.15 & 2.66 & \textbf{1.80} \\
Median position  & \textbf{1.00} & \textbf{1.00} & \textbf{1.00} \\
Min position     & \textbf{1.00} & \textbf{1.00} & \textbf{1.00} \\
Max position     & \textbf{17.0} & 30.0 & \textbf{17.0} \\
Stdev position           & 3.03 & 5.67 & \textbf{2.72} \\
Hit@K = 1        & 0.707 & 0.732 & \textbf{0.756} \\
Hit@K = 3        & 0.902 & 0.927 & \textbf{0.951} \\
Hit@K = 5        & 0.927 & 0.927 & \textbf{0.951} \\
Hit@K = 10       & 0.951 & 0.951 & \textbf{0.976} \\
\hline
\end{tabular}
\vspace{-5pt}
\caption{Disease-only Query Evaluation}
\label{table:disease-only}
\end{table}
\vspace{-15pt}
Results for the quantitative evaluation are contained in Tables \ref{table:disease-modality} and \ref{table:disease-only}. The embedding-based score had a lower average matching position than BM25 for both the disease+modality and disease-only queries, with the full hybrid search achieving the best results. High hit rates were observed, including over 95\% at a Hit@K = 3 with the hybrid model.

We additionally quantified inference time of the search algorithm using the simulated query set generated during the hyperparameter optimization. Across 22,552 queries with 5.8 average words per query, the mean inference time for the hybrid search was 0.38 seconds (SD of 0.11 seconds).

\subsection{Qualitative Examples}
\begin{figure}[htbp] 
\centering
\includegraphics[width=1.0\linewidth]
{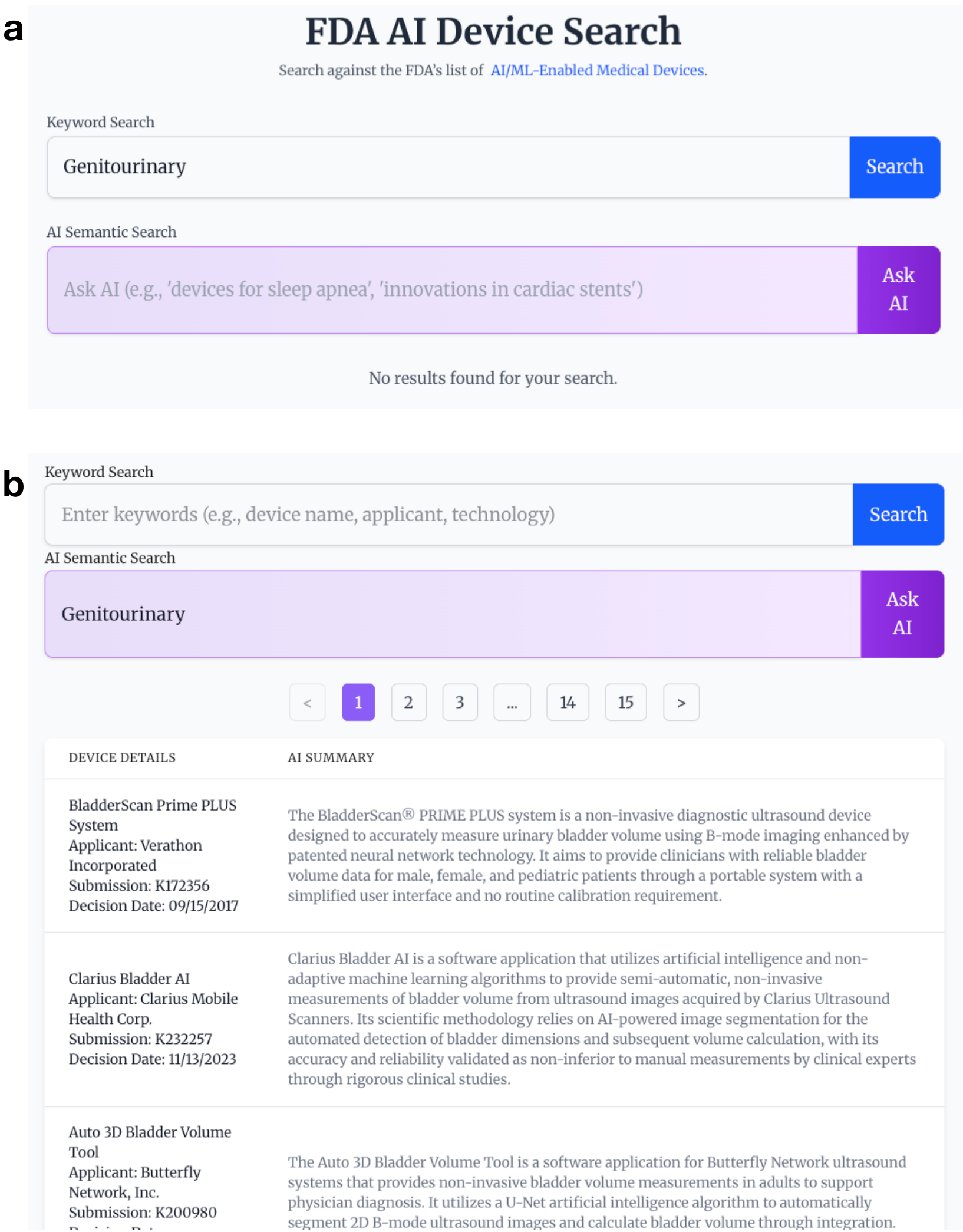}
\caption{Example illustrating benefits of semantic search. a) Keyword search finds no matches for ``Genitourinary''. b) Semantic search finds relevant devices.} \label{fig:keyword_semantic_comparison}
\end{figure}

Fig. \ref{fig:sample-ui} illustrates the website UI and how the basic keyword search is useful for targeted queries like ``breast density''. Fig. \ref{fig:keyword_semantic_comparison} illustrates the utility of the AI-enabled semantic search: while the keyword search for ``genitourinary'' returns no matches, the semantic search identifies relevant devices. Additional examples are contained in Appendix \ref{additional_examples}.

\vspace{-5pt}
\section{Conclusion}
We developed \url{fda-ai-search.com} to enable efficient searching of FDA-authorized AI-enabled medical devices. We anticipate that it will assist healthcare providers in identifying devices aligned with their clinical needs and support developers in formulating novel AI applications. Given the current lack of curated information, quantitative evaluation is challenging, but our pilot study supports the efficacy of the retrieval algorithm. We note that the website is intended as an assistive tool and may make errors, including the potential for biases or hallucinations inherent to LLM-based feature extraction, for which future user studies will be important. We plan on updating the website in step with the FDA’s AI device list and aim to iteratively improve the system based on user feedback.

\acks{W.L. acknowledges funding support from NIBIB award R21EB035247 and NLM award R01LM014775.}

\bibliography{main}

\clearpage
\appendix

\section{Method Details}\label{apd:first}

\subsection{Feature Extraction}

When extracting features from the summary PDFs, the PDF was first split into chunks of 200 pages or less. The vast majority ($\sim$99\%) of PDFs were less than 200 pages. Each chunk was then first summarized by Gemini-2.5-flash using the following prompt:

\begin{quote}
Summarize the following chunk of a document in two paragraphs. Do NOT include anything about the document type (e.g., FDA 510(k) letter)
\end{quote}

If there was more than one chunk, the model was queried to create an aggregate summary across the chunk summaries, otherwise the chunk summary was used as the aggregate summary. The following prompt was then used to extract the features from the aggregate summary:

\begin{quote}
Analyze the following summary of a document and provide:
1. A thorough two-paragraph summary that distills all relevant content about the device's purpose, methology, science, and results from the entire document.
2. Exactly 10 salient keywords from the entire document.
3. Exactly 5 insightful questions that a clinician or scientist might ask about the entire document to yield good results for further investigation or understanding.
4. Based on the keywords and generated questions, a list of 5 concepts pertaining to the entire document.
5. Based on all of the above, a 2 sentence thesis, a clear statement of purpose, methology, science of the device for the entire document.
Summary content:
\end{quote}

The prompt used to generate the `query match' features was as follows:
\begin{quote}
Based on the following data for a medical device, please generate three distinct search queries that a clinician would use to find this device. The queries should be specific and relevant to the device's characteristics and intended use.
\end{quote}
The data provided for this prompt consisted of a json of the original five extracted features.

\subsection{Hyperparameter Optimization}
\label{hyperopt_appendix}

The prompt for creating the simulated queries was as follows:

\begin{quote}
You are an expert medical researcher.
Based on the following thesis and key concepts from a medical device's FDA summary,
generate a concise and clinically relevant search query that a clinician or researcher might use to find information about similar devices or technologies. Do NOT include anything about AI, ML, or what those mean.
Only return the 1-3 word medical search query. Return only the query itself, without any preamble or explanation..
Thesis: ``\{thesis\}"
Key Concepts: ``\{concepts\}"
Clinically Relevant Search Query:''
\end{quote}

The final set of embedding weights ($w_i$) resulting from the hyperparameter search was as follows:
\begin{lstlisting}[language=Python]
{
    'keywords': 0.134207,
    'questions': 0.226103,
    'thesis': 0.094972,
    'search_boost': 0.029563,
    'query_match_1': 0.217395,
    'query_match_2': 0.241111,
    'query_match_3': 0.056650,
}
\end{lstlisting}

The final value of $\lambda$ was 0.8.

\subsection{Additional Qualitative Examples}
\label{additional_examples}

\begin{figure}[htbp]
  \centering
\includegraphics[width=1.0\linewidth]{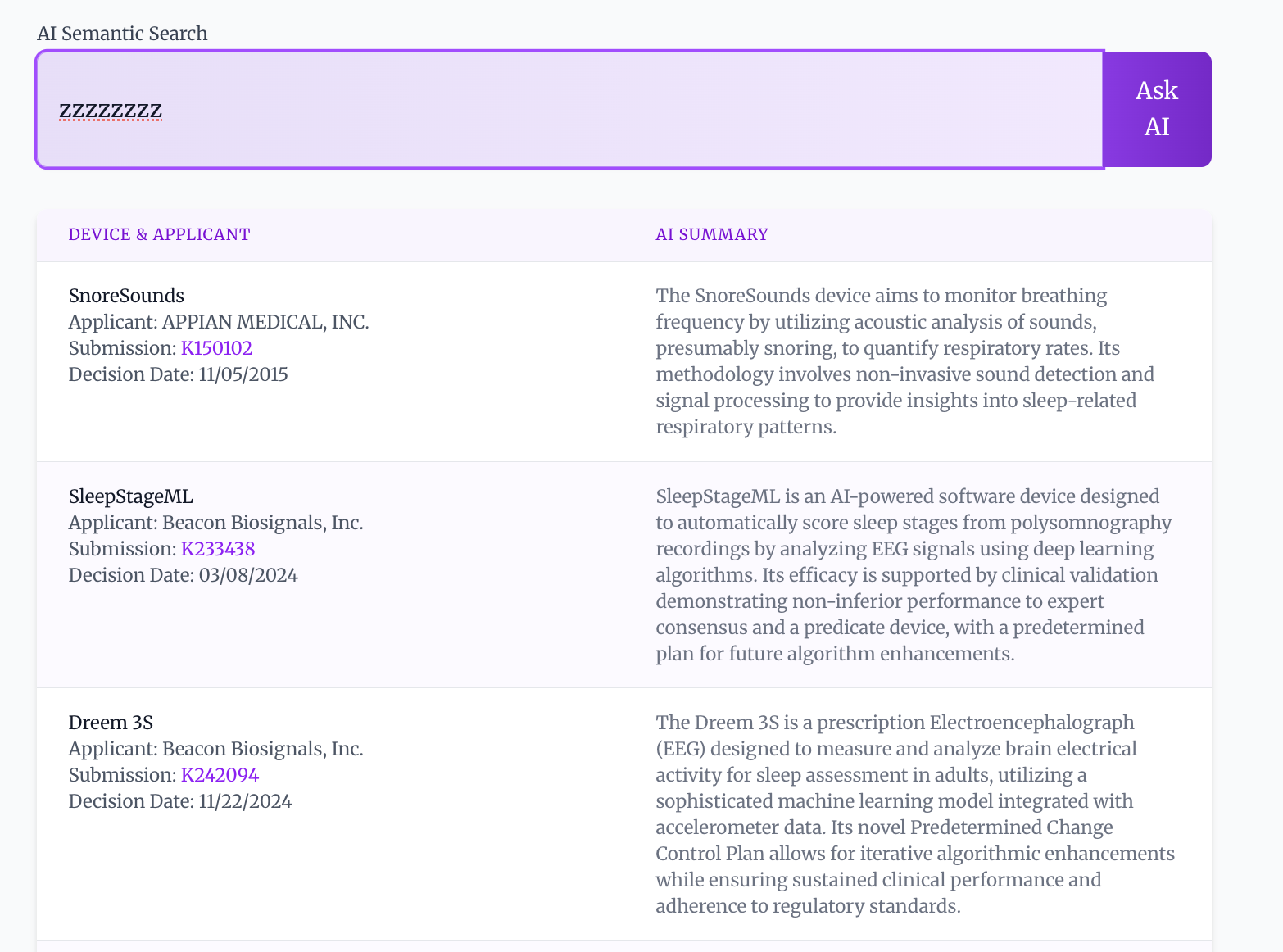}
  \caption{Semantic search results for ``zzzzzzzz''.}
  \label{fig:zzz}
\end{figure}

\begin{figure}[htbp]
  \centering
\includegraphics[width=1.0\linewidth]{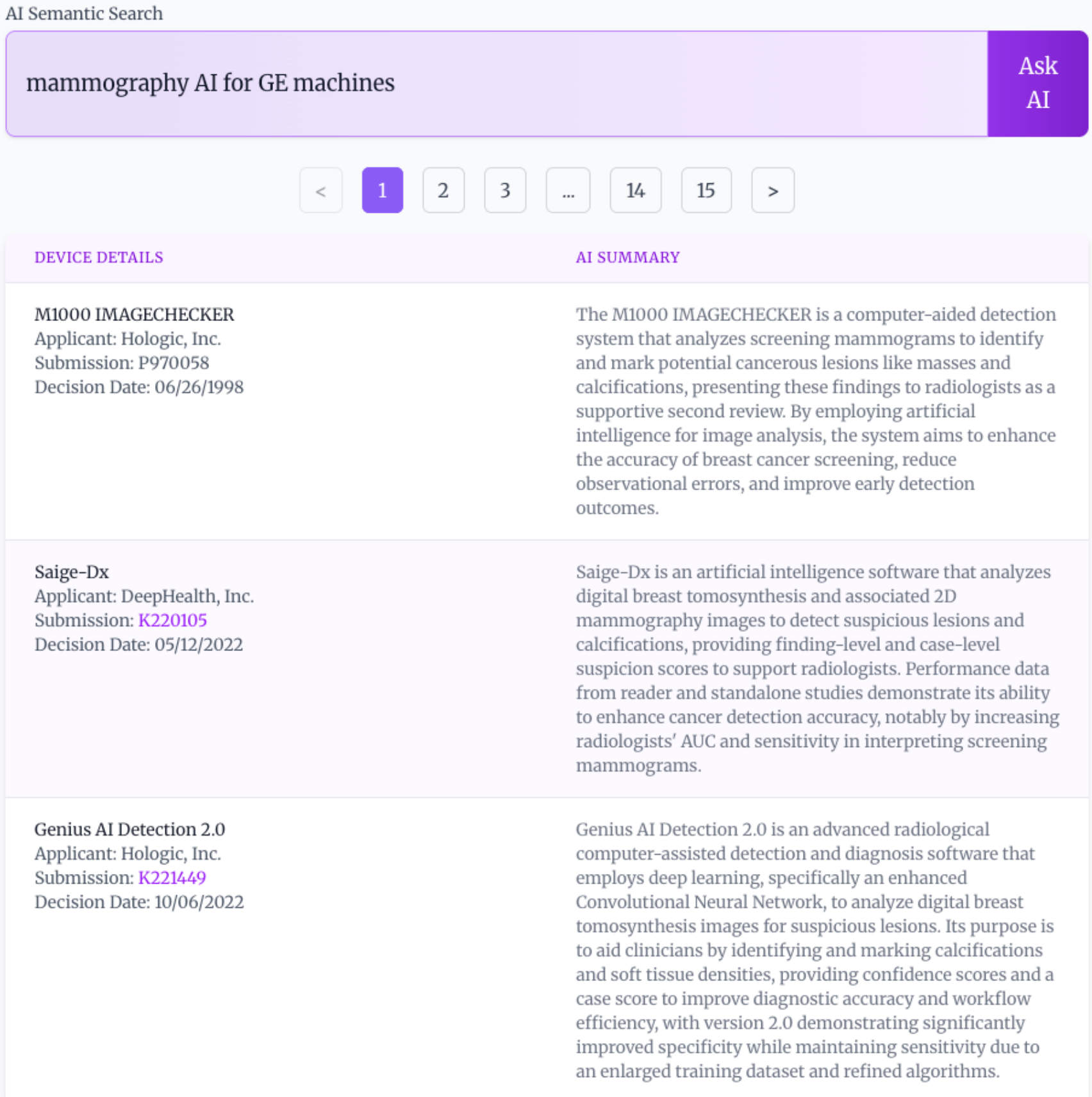}
  \caption{Failure mode example where ``mammography AI for GE machines'' retrieves AI devices for mammography in general, rather than those specifically compatible with GE machines.}
  \label{fig:failure}
\end{figure}

An extreme example of semantic search is illustrated in Fig.~\ref{fig:zzz}, where the query ``zzzzzzzz'' retrieves devices related to sleep. An example downside of semantic search is illustrated in Fig.~\ref{fig:failure}, where ``mammography AI for GE machines'' retrieves devices using AI for mammography, but closer inspection reveals that not all of the devices listed are compatible with GE machines.

\end{document}